\documentstyle[preprint,aps]{revtex}
\draft
\begin{document}
\bibliographystyle{plain}
\newcommand{\be}{\begin{equation}}
\newcommand{\ee}{\end{equation}}
\newcommand{\bea}{\begin{eqnarray}}
\newcommand{\eea}{\end{eqnarray}}
\newcommand{\ddl}[1]{\stackrel{\leftarrow}{\partial\over\partial#1}}
\newcommand{\ddr}[1]{\stackrel{\rightarrow}{\partial\over\partial #1}}
\newcommand{\ddd}[2]{{\partial^2\over\partial #1\partial #2}}

\title
{Small Disks and Semiclassical Resonances}

\author{P. Rosenqvist and N. D. Whelan\cite{newadd}}
\address{Centre for Chaos and Turbulence Studies,
Niels Bohr Institute,
Blegdamsvej 17, DK-2100, Copenhagen \O, Denmark}

\author{A. Wirzba}
\address{Institut f\"{u}r Kernphysik, Technische Hochschule Darmstadt,
Schlo{\ss}gartenstr.9, 64289 Darmstadt, Germany}

\date{\today}

\maketitle

\begin{abstract}
We study the effect on quantum spectra of the existence of small
circular disks in a billiard system. 
In the limit where the disk radii vanish there
is no effect, however this limit is approached very slowly so that even
very small radii have comparatively large effects.  We include diffractive
orbits 
which scatter off the small disks in the periodic orbit expansion.
This situation is formally similar to edge diffraction except that
the disk radii introduce a length scale in the problem such that for
wave lengths smaller than the order of the disk radius we recover the 
usual semi-classical
approximation; however, for wave lengths larger than the order of the
disk radius there is
a qualitatively different behaviour.
We test the theory by successfully estimating the positions of scattering
resonances in geometries consisting of three and four small disks.
\end{abstract}


The presence of discontinuities in classical Hamiltonian systems implies
the necessity of a closer study
of the quantum mechanics when doing semiclassical periodic orbit theory 
\cite{gutz} and has been the theme of numerous recent papers
\cite{creep,vwr1,vwr2,ndw1,ps,smil,ndw2,bw}.
The approach is to study the quantum
scattering problem near the discontinuity, combine this with classical
information about classical trajectories 
away from the discontinuity to find global quantities such as the trace of the 
quantum Green function. In doing so, we
maintain the local-global duality inherent in periodic orbit theory.
In this paper we discuss one class of discontinuity, that of small circular
scatterers.
In the context of the Sinai billiard, the perturbative effect of a small disk
in the quantum \cite{berry} and classical problems \cite{fried,dahlqvist}
was studied, but not with a scattering interpretation.
By small, we mean smaller than the typical wavelength in the
problem. The opposite limit, of scatterers much larger than 
a typical wavelength, can be evaluated using classical periodic orbits 
reflecting off the disk plus
creeping diffraction to account for the discontinuity associated with glancing
orbits \cite{creep,vwr1,vwr2,smil}.

We will analyse billiard systems in two dimensions and therefore seek the
Green function of the Helmholtz equation
\be \label{jesse}
\left(\nabla^2+k^2\right)\psi = 0
\ee
with some specified boundary conditions. In the absence of any boundaries, the
Green function between a source at $x'$ and a receiver at $x$ is
\bea 
G_f(x,x',k) & =       & -{i\over 4}H_0^{(+)}(k|x-x'|) \nonumber \\
            & \approx & {1\over\sqrt{8\pi k|x-x'|}} e^{i(k|x-x'|-3\pi/4)}.
\label{freespace}
\eea
The second line, which is asymptotic in $k|x-x'|$, will be useful later in
the discussion.
In the exterior of a disk of radius $a$ with Dirichlet boundary conditions, 
the first line of (\ref{freespace}) is modified to
\bea
G(x,x',k) & = & -{i\over 8} \sum_{m=-\infty}^\infty e^{im(\theta-\theta')}
H_m^{(+)}(kx')\left(H_m^{(-)}(kx)-{H_m^{(-)}(ka)\over H_m^{(+)}(ka)}
H_m^{(+)}(kx)\right) \;\;\;\;\; x'\geq x \nonumber \\
& = & G_f(x,x',k) + {i\over 4}\sum_{m=-\infty}^\infty e^{im(\theta-\theta')}
{J_m(ka)\over H_m^{(+)}(ka)}H_m^{(+)}(kx')H_m^{(+)}(kx), \label{diskgf}
\eea
where $\theta$ and $\theta'$ are the polar angles of points $x$ and $x'$ as
measured from the disk centre. 
The first line follows from using Graf's addition formula for 
$H_0^{(+)}(z)$ \cite{as} together with the S-matrix of the disk 
scattering problem. The second line can be seen to equal the first by 
another application of Graf's addition formula and the expansion of $J_m(z)$
in terms of $H_m^{(\pm)}(z)$. 
If $x>x'$, one must interchange $x$ with $x'$ 
and $\theta$ with $\theta'$ in the first line of Eq.~(\ref{diskgf}) but
the second line rests unchanged.
This has the appealing structure of being
the free space Green function plus a correction. Whether the correction is
small or large depends on the geometry of the problem and on the wave number
$k$.

We now make the assumption that the disk radius $a$ is much smaller than the
typical distance to points $x$ and $x'$ from the disk. We further assume the
semiclassical condition $kx,kx'\gg 1$. Recalling the asymptotic relation
$H_m^{(+)}(z)\approx \exp(-im\pi/2)H_0^{(+)}(z)$ (assuming $|z|\gg 1$ and 
$|z|>m$), we obtain
\bea
\lefteqn{\sum_{m=-\infty}^\infty e^{im(\theta-\theta')}
{J_m(ka)\over H_m^{(+)}(ka)}H_m^{(+)}(kx')H_m^{(+)}(kx)\approx} 
\;\;\;\;\;\;
\;\;\;\;\;\;
\;\;\;\;\;\;
\;\;\;\;\;\;
\;\;\;\;\;\;
\;\;\;\;\;\;
\;\;\;\;\;\;
\nonumber \\
& & H_0^{(+)}(kx')H_0^{(+)}(kx)
\sum_{m=-\infty}^\infty e^{im(\theta-\theta'-\pi)}{J_m(ka)\over H_m^{(+)}(ka)}.
\label{beingthere}
\eea
Although the asymptotic forms used for $H_m^{(+)}(z)$ breaks down for
$m>|z|$, this is not important in Eq.~(\ref{beingthere}) since the condition
$x,x'\gg a$ implies that the breakdown begins 
for values of $m$ such that the factor $J_m(ka)/H_m^{(+)}(ka)$ is already
very small. We then conclude that in the presence of a small disk centred 
at position $\xi$,
the Green function between two points separated by an angle $\phi$ as 
measured from $\xi$ is approximately
\be \label{casablanca}
G(x,x',k) \approx G_f(x,x',k) + d(\phi)G_f(x,\xi,k)G_f(\xi,x',k)
\ee
where we have defined a diffraction constant
\be \label{bluevelvet}
d(\phi) = -4i\sum_{m=-\infty}^\infty e^{im(\phi-\pi)}
{J_m(ka)\over H_m^{(+)}(ka)}.
\ee
Using the relations 
$J_{-m}(z)=(-1)^mJ_m(z)$ and $H_{-m}^{(+)}(z)=(-1)^mH_m^{(+)}(z)$, we can
replace the exponentials in this expression by cosines thereby
displaying the time-reversal property $d(\phi)=d(-\phi)$. Furthermore, this
sum converges since the ratio $J_m(ka)/H_m^{(+)}(ka)$ decreases factorially
with $m$ for $m>|ka|$.
In the more general case of a non-circular but small scatterer,
we must insert the full $S$ matrix into Eq.~(\ref{diskgf}).
The analysis remains largely unchanged; in particular
Eq.~(\ref{casablanca}) still applies however the diffraction constant 
is then a function of both angles and not just their difference. We
then obtain
\be \label{gendisk}
d(\theta,\theta') = 2\sum_{m,m'=-\infty}^{\infty}
e^{i(m'\theta-m\theta'-(m+m')\pi/2)} T_{mm'},
\ee
where we have defined the $S$ matrix through
$S_{mm'}=\delta_{mm'}-iT_{mm'}$.

The Green function (\ref{casablanca}) has a direct contribution
as if there were no disk plus a contribution 
in the form of a product of Green functions which arises from scattering
off the disk. This is the same structure which exists in the presence
of vertices \cite{kel} where
we obtain the diffraction constant from the solution
of the wedge scattering problem solved by Sommerfeld \cite{somm}.
Despite the similar form, there are two aspects of the problems which
are quite different.
The small disk diffractor has no internal orientation but does have an
internal length scale, $a$. In contrast, a wedge has an internal orientation,
as given by the direction of its normal, but has no internal length scale.
The existence of an orientation means that
there are choices of incoming and outgoing angles for which the vertex
diffraction constant diverges whereas this never happens for the disk. On
the other hand, the lack of an internal length scale means that the vertex
diffraction constant is independent of $k$ whereas for the disk it is
clearly $k$-dependent. The systems do share the property that we can trivially
extend them to include problems with a potential $V(x)$ \cite{bw}.
We assume that the potential does not change much in a wavelength and thereby
compute the diffraction constant with $k=\sqrt{2m(E-V(\xi))}/\hbar$, the 
wave number at the disk. Eqs.~(\ref{casablanca}) and 
(\ref{bluevelvet}) then apply
where $G_f$ is the Van-Vleck approximation to the Green function for that 
potential in the absence of a disk.

So far we have made no assumption on the value of $ka$.  For $ka\gg 1$ we 
recover the expected geometrical structure, as we discuss below.
There is qualitatively
different behaviour for $ka\ll 1$ and a cross-over for $ka\sim 1$.
In the limit $ka\rightarrow 0$ we note that
$J_m(ka)/H_m^{(+)}(ka) \approx i\pi(ka/2)^{2m}/m!(m-1)!$ for $m\neq 0$ and
only the $m=0$ term contributes significantly. We call this the
s-wave limit. {}From the approximations
\be \label{taxidriver}
J_0(ka) \approx 1
\;\;\;\;\;\;\;\;\;
Y_0(ka) \approx {2\over\pi}\left(\log\left({ka\over 2}\right)+\gamma_e\right)
\ee
valid for small $ka$
(where $\gamma_e=0.577\ldots $ is Euler's constant), we  derive the s-wave
approximation
\be \label{bluelagoon}
d \approx {2\pi \over \log\left({2 \over ka}\right) - \gamma_e + 
i{\pi\over 2}},
\ee
which is independent of scattering angle $\phi$.
As $ka\rightarrow 0$, the denominator of Eq.~(\ref{bluelagoon})
grows logarithmically so that the diffraction constant goes to zero and the
disk has no effect, which is reasonable.
However, this happens very slowly so that even for very small values of $ka$
there is still an appreciable effect, as we will demonstrate.

It might seem surprising that the diffraction constant vanishes
as $a\rightarrow 0$ since we are demanding that the wave function vanish at
a point, and it might be thought that this should have some effect. That
this is not so can be understood with the example of an annulus in which the
central disk is very small. Although the eigenfunction does indeed vanish
on the disk, it increases very rapidly so that within a small distance
the eigenfunction is indistinguishable from one in which there were no central
disk. In this sense, the wavefunctions (and eigenvalues) are virtually
indistinguishable from those corresponding to the disk-free system.
In Ref.~\cite{seba}, the author argues
that disks of zero radius continue to have an effect, but the system he was
considering was equivalent to an infinitely thin line charge in an
electro-magnetic wave guide (see also Ref.~\cite{seba2} for the three
dimensional generalisation.) In his language, our disk is uncharged and
there is no contradiction between his conclusions and ours.
The difference is in the order one takes the limits $a\rightarrow 0$ and
$k\rightarrow\infty$. In Ref.~\cite{seba}, one starts with the first limit
(while maintaining a finite interaction) whereas we consider the second limit
while holding $a$ small but fixed. As a result, the problems are
quite distinct. For example, in the short wavelength limit the disks considered
here will start having a large, classical effect (i.e. when $ka$ is of order 1)
whereas in the system mentioned above the effect of the 
scatterer vanishes for large $k$ \cite{cheon}.

In a system composed of several scattering centres, the Green function between
points labelled $x'$ and $x$ will receive contributions from paths
which diffract several times. The contribution of one such path,
labelled $p$, is a
simple generalisation of the second term of (\ref{casablanca}) and equals
\be \label{general}
G_p(x,x',k) \approx G_f(x,\xi_n,k)d_n
\left\{\prod_{j=1}^{n-1} G_f(\xi_{j+1},\xi_j,k)d_j\right\}G_f(\xi_1,x',k).
\ee
The quantities $\xi_j$ indicate the scattering points - in this case
the centres of the disks with the subscripts $j$ indicating the order in which
they are encountered for that path. The path is
composed of $n$ scatterers with a diffraction constant $d_j$
associated with each one. The details of the trace integral are worked out
in Refs.~\cite{vwr1,ps,bw}; here we simply sketch the derivation. 
We stress that all of these calculations are to leading order in
$\hbar$ (or $1/k$ in this context.)
The criterion of stationary phase selects periodic orbits
which are everywhere classical except at the singularities 
where they diffract by an arbitrary angle. A periodic orbit is composed of
$n$ classical segments connected by $n$ diffractions.

If we identify $x$ with $x'$ in (\ref{general}) we see that it is a
closed cycle of Green functions in which the segment between the scatterers
$\xi_n$ and $\xi_1$ is ``cut" by the point $x$ which lies between them.
In general, the trace integral associated with a periodic orbit
labelled $\gamma$ will be evaluated by integrating over
all choices of $x$. This means that we must allow
$x$ to cut open the cycle of Green's functions between any two consecutive 
scatterers - each possibility physically
corresponds to $x$ lying between
that pair of scatterers. For
this purpose, we define a parallel coordinate $z$ which runs along the
periodic orbit from $\xi_n$ to $\xi_1$ to $\xi_2$ etc. until it returns to
$\xi_n$. 
When $z$ is between $\xi_i$ and $\xi_{i+1}$, the point $x$ is between
these two scatterers (where we identify $n+1$ with 1.) 
At each point along the orbit, we 
define a transverse coordinate $y$ so that $x$ is parameterised by the pair
$(y,z)$, as for geometric orbits \cite{gutz}. The
integrand of the trace integral
is similar to (\ref{general}) but where we cut the cycle of Green's functions
between $\xi_i$ and
$\xi_{i+1}$ (as governed by $z$) rather than between $\xi_n$ and $\xi_1$ so
that the trace integral associated with $\gamma$ is
\be \label{traceint}
\oint dz\int dy G_\gamma(x,x,k) 
= \sum_{i=1}^n\int_{\xi_i}^{\xi_{i+1}}dz\int_{-\infty}
^\infty dy G_f(x,\xi_i,k)d_i
\left\{\prod_{j\neq i} G_f(\xi_{j+1},\xi_j,k)d_j\right\}
G_f(\xi_{i+1},x,k).
\ee
Due to stationary phase, only
those points close to the orbit contribute significantly so that
to leading order, all the diffraction constants $d_j$ are
independent of $x$ and can be considered invariant properties
of the periodic orbit. Therefore, the only $x$ dependence is in the
two Green functions connecting $x$ to the adjacent scatterers; 
all other factors in the integrand are constant.
For each value of $z$, we evaluate the transverse integral $\int dy$
by stationary phase; this yields a factor which is independent of $z$ and is
proportional to the Green function between the two adjacent scatterers.
Combining this with all the constant factors in the integrand gives the
product of the closed cycle of Green functions and diffraction constants
from one scatterer to the next, $\prod_{j=1}^nG_f(\xi_{j+1},\xi_j,k)d_j$.
We get this same invariant factor after doing the
$y$ integral at any point $z$ along any segment of the periodic orbit so 
there is no explicit $z$ dependence remaining.
The integral parallel to the orbit $\oint dz=L_\gamma$ is then simply the
length of the periodic orbit, as also happens for geometric orbits.
The final result is
\bea
g_\gamma(k) & \approx &
-i{L_\gamma\over 2k} \left\{\prod_{j=1}^{n_\gamma} G_f(\xi_{j+1},\xi_j)d_j
\right\}\nonumber\\
& \approx & -i{L_\gamma\over 2k}
\left\{\prod_{j=1}^{n_\gamma}{d_j\over\sqrt{8\pi kL_j}}\right\}_\gamma
\exp\{i(kL_\gamma-3n_\gamma\pi/4)\}. \label{thebicyclethief}
\eea
We  have made use of (\ref{freespace}) where $L_j=|\xi_{j+1}-\xi_j|$.
This expression involves one fewer Green function than (\ref{general}).
(The additional Green function contributes a constant proportional to 
$e^{-i3\pi/4}/\sqrt{k}$ which combines with a factor proportional to 
$e^{i\pi/4}/\sqrt{k}$ from the stationary phase
integral to give the prefactor of $-i/2k$.) 
Eq.~(\ref{thebicyclethief}) is the contribution of a
single diffractive periodic orbit, in general we must sum over all
such orbits as well as over all purely geometric orbits to get the
total trace. For this reason, we have introduced the subscript
$\gamma$ on the index $n_\gamma$ in the above equation.

Following Refs.~\cite{vwr1,vwr2,ndw2} we write down the semiclassical 
diffractive zeta function \cite{zet} whose zeros approximate the exact
quantum resonances,
\be \label{bettyblue}
\zeta^{-1}_{\mbox{diff}} = \prod_\gamma(1-t_\gamma)
\ee
where
\be \label{bleu}
t_\gamma = \left\{\prod_{j=1}^n{d_j\over\sqrt{8\pi kL_j}}\right\}_\gamma
\exp\{i(kL_\gamma-3n_\gamma\pi/4)\}
\ee
This results follows from the semiclassical approximation
\be \label{derv}
{dt_\gamma \over dk^2} \approx {iL_\gamma \over 2k}t_\gamma
\ee
so that the sum over all diffractive orbits in Eq.~(\ref{thebicyclethief}) is
the logarithmic derivative of the zeta function (\ref{bettyblue}). (We
take the derivative with respect to $k^2$ since the trace of the 
Green function (\ref{thebicyclethief}) 
is properly thought of as a function of $k^2$ and not of $k$.)
The product is over just the primitive orbits; their repeats have already
been summed. In a system with coexisting geometric and diffractive orbits,
we need to multiply the corresponding zeta functions \cite{vwr1,vwr2}. 
The result is a purely formal product which must be regulated differently
for scattering \cite{ce,aac1,aac2} and 
bound \cite{berkeat} problems so that its 
zeros are the semiclassical eigenvalues of the full problem and not the zeros
of the individual terms in the product. The diffractive zeta function
(\ref{bettyblue}) 
involves no additional product as happens for geometric orbits \cite{vor},
resulting in there being only leading resonances in scattering calculations
\cite{ndw1,ndw2}.

We now specialise the discussion to scattering geometries featuring
three and four small disks arranged symmetrically in the plane
\cite{eck,gr_cl,gr_sc,gr_qu,ce1,ce2,pinball,fd}.
We first discuss the three disk
problem as shown in Fig.~\ref{mark_3}a (we exaggerate the size of the disks to
make the discussion clearer.) Starting from one of the disks,
there are two
distinct processes. We can go to one of the other two disks and 
either scatter back to the original disk or scatter on to the third disk.
We assign these two processes the symbols 0 and 1 respectively \cite{ce1}.
The weights are $t_0=d(0)u$ and $t_1=d(\pi/3)u$ where the factor
\be \label{liarsmoon}
u = {1\over \sqrt{8\pi kR}}\exp\{i(kR-3\pi/4)\}
\ee
is common to both orbits and $R\gg a$ is the inter-disk spacing. 

Notice that
there is a $C_{3v}$ symmetry to this problem \cite{gr_sc,gr_qu,ce1,ce2}
consisting of the identity, rotations by $\pm 2\pi/3$ and reflections through
the three symmetry axes. This group has three irreducible representations which
are called $A_1$, $A_2$ and $E$.
We can make use of this symmetry by considering dynamics
in the fundamental domain \cite{robbins},
which in this case is a wedge consisting of one sixth of the plane as
indicated in Fig.~\ref{mark_3}a. One does this by following a trajectory
and using the symmetry operations to map the trajectory back into the 
fundamental domain whenever it crosses a boundary. In the fundamental 
domain of the three disk problem, there is only one half disk.
A trajectory can leave this half disk in only one direction,
which is labelled A. Upon encountering the border of the fundamental domain,
a reflection operation is applied so that the trajectory returns to the disk,
where it
has two choices. It can either diffract back onto A or it can diffract into
the direction A'. In the second case, we apply a reflection operator again
to map this back onto A. These two possibilities are both diffractive
periodic orbits of the fundamental
domain and have the weights $t_0$ and $t_1$ discussed above. Each orbit
has an additional
group theoretic weight given by the characters (in the representation being
considered) of the group operations
needed to keep the orbit in the fundamental domain \cite{ce2}.

In general there are longer periodic orbits as labelled by whether they
back scatter or forward scatter at each encounter with the disk. These can
be then labelled by a binary sequence of 0's and 1's. However, there is
a multiplicative property to the weights such that the weight of any long
orbit is equal to the product of the weights of shorter cycles.
For example $t_{001}=t_0^2t_1$ since they both equal $d_0^2d_1u^3$.
This property means that we can represent the zeta function as being the
determinant of a Markov graph \cite{mark}, 
which is drawn in Fig.~\ref{mark_3}b. The single
node in the graph, A, is connected to itself by the processes 0 and 1
described above.

All the characters of the totally symmetric representation $A_1$ are unity,
which simplifies its discussion. To find its zeta function, we simply read
off from the Markov graph all nonintersecting closed loops. In this case there
are only two and we get the simple result
\be \label{a_1--3d}
\zeta^{-1}_{A_1} = 1-t_0-t_1.
\ee
This formula agrees with the result found in Refs.\cite{ce1,ce2} for the
special case where all the higher order ``curvature corrections'' 
\cite{ce,aac1,aac2} vanish identically. This vanishing is simply a result
of the fact that we have a one node graph so we only need consider weights
of topological length one. Armed with this rule, we can then read off from
Refs.\cite{ce1,ce2} the zeta functions of the other two
representations. These are
\be \label{loveanddeath}
\zeta^{-1}_{A_2} = 1+t_0-t_1 \;\;\;\;\;\;\;\;\;\;\;\;\;\;\;\;\;\;\;
\zeta^{-1}_{E} = 1+t_1+t_1^2-t_0^2.
\ee
We could have ignored the symmetry decomposition and simply drawn the
six node Markov graph of the full problem as shown in Fig.~\ref{mark_3a}. 
The vastly increased number
of closed loops in comparison to Fig.~\ref{mark_3} underlines the 
advantage of using the symmetry reduction. The rule for finding the zeta
function is to find all non-intersecting closed loops and products of
non-intersecting closed loops. A product of $n$ non-intersecting 
closed loops has a relative sign $(-1)^n$.
Carefully enumerating all such loops of the full graph, we find its
zeta function to be
\be\label{smoke}
\zeta^{-1} = 1 - 3t_0^2 - 2t_1^3 + 3t_0^4 - 3t_0^2t_1^2 - t_0^6 + t_1^6 -
3t_0^2t_1^4 + 3t_0^4t_1^2.
\ee
This equals the product $\zeta^{-1}_{A_1}\zeta^{-1}_{A_2}\zeta^{-2}_E$ 
of the symmetry decomposed zeta functions
above. In addition to the additional complexity of its Markov graph and
zeta function, the full zeta function has the further
disadvantage that we do not know to which symmetry class one of its zeros 
belongs. However, this exercise is useful in verifying that our use of 
the results of Ref.~\cite{ce1,ce2} is well founded.

The exact resonances of this geometry can be found numerically by finding
the zeros of the determinant of a matrix.  This matrix is \cite{gr_qu}
\be \label{matm}
~M_{nm} = \delta_{nm} + A_{nm}
\ee
where for the $A_1$ resonances
\be\label{mata}
A_{nm} = {J_n(ka) \over H_m^{(+)}(ka)}
\left[\cos\left({\pi\over 6}(5n-m)\right)
H_{n-m}^{(+)}(kR) + (-)^m\cos\left({\pi\over 6}(5n+m)\right)H_{n+m}^{(+)}(kR)
\right].
\ee
Expressing $\mbox{det} M$ in a cumulant expansion \cite{creep,trcl}, 
valid because $A$ is trace-class \cite{trcl}, yields
\be \label{cumexp}
\mbox{det} M = 1 + \mbox{tr} A - {1\over 2}\left(\mbox{tr}A^2 - 
\left(\mbox{tr}A\right)^2\right) + \cdots
\ee
where from Eq.(\ref{mata}) one obtains \cite{creep}
\be \label{tra}
\mbox{tr} A = \sum_{m=-\infty}^\infty {J_m(ka) \over H_m^{(+)}(ka)}\left(
\cos\left({2\pi m\over 3}\right)H_0^{(+)}(kR) + H_{2m}^{(+)}(kR)\right).
\ee
We now impose the same constraints as before, namely $kR\gg 1$ and $R\gg a$ so
that we can replace $H_{2m}^{(+)}(kR)$ by $\cos(m\pi)H_0^{(+)}(kR)$ and
\be \label{trasimp}
\mbox{tr} A \approx H_0^{(+)}(kR) \sum_{m=-\infty}^\infty
{J_m(ka) \over H_m^{(+)}(ka)} 
\left(\cos(2\pi m/3) + \cos(m\pi)\right).
\ee
Then using
Eqs.(\ref{bluevelvet}) and (\ref{liarsmoon}) and the asymptotic form of
$H_0^{(+)}(kR)$, we find
\be \label{trasimp2}
\mbox{tr} A \approx -u\left(d(0) + d(\pi/3)\right) = -(t_0+t_1).
\ee
We see that truncating the cumulant expansion at the term linear in $A$ and 
invoking the relevant approximations gives
the same equation for $\mbox{det} M = 0$ as we earlier derived for
$\zeta^{-1}=0$.  This is reassuring since it means that we understand the
error caused by replacing $H_m^{(+)}(z)$ by $\exp{(-im\pi/2)}H_0^{(+)}(z)$ in 
Eqs.(\ref{beingthere}) and (\ref{trasimp});
it is the same as neglecting higher order terms in the
cumulant expansion. As shown in Refs.~\cite{creep,trcl}, this is equivalent to
neglecting higher order curvature corrections in the cycle expansion.
The fact that the semiclassical approximation can be made on the
level of the traces of the scattering kernel $\mbox{tr}A^n$ 
(\ref{mata}) which result from the defining cumulant expansion
(\ref{cumexp}) provides an alternate method to arrive directly at the
zeta function $\zeta^{-1}_{\mbox{diff}}$. This method
does not require
closed expressions for the trace of the Green function, does not invoke the
semiclassical relation (\ref{derv}) and, most importantly,
appears in a curvature-regulated form
\cite{creep,trcl}. In particular, one can use this to read the weights
$t_\gamma$ (\ref{bleu}) 
directly from the Green's function product 
$\prod_{j=1}^nG_f(\xi_{j+1},\xi_j,k)d_j$, which is just the closed path 
equivalent of the open path Green's 
function (\ref{general}).

The identification between the quantisation conditions 
$\zeta^{-1}_{A_1}=0$ and $\mbox{det} M = 0$ tells us something else.
In Ref.~\cite{creep} it is shown that one can extract the contribution of
geometric orbits and diffractive creeping orbits from $\mbox{tr} A$ by
invoking Watson contour integration to replace the sum of Eq.(\ref{tra}).
This means that the diffraction constant contains information about periodic
orbits and creeping. Therefore, even in the limit $ka\gg 1$, the formalism
described here still applies, the price being the necessity to include many
terms in calculating the diffraction constant (\ref{bluevelvet}). We therefore
have a uniform picture. For large values of $ka$, one invokes geometric and
creeping orbits but for intermediate and small values one invokes the small
disk
scattering theory elucidated here. These are guaranteed to match smoothly.
Although this was shown explicitly only for two and three disk systems,
the same will hold for any number of disks in any geometrical arrangement.

We show the exact and semiclassical results in Fig.~\ref{3d_res} 
for the $A_1$ and $E$
resonances together with the approximations using geometric orbits. The
resonances are shown in the complex $k$ plane and are
measured in units of $1/R$. In Fig.~\ref{3d_res}a
we show the results for the $A_1$ resonances for $R/a=60$
so that the cross-over condition 
${\mbox{Real}}\{ka\}=1$ corresponds to ${\mbox{Real}}\{kR\}=60$. The
minimum, which is developing at the right of the figure, has a geometrical
interpretation in terms of interference between the two shortest geometrical
orbits in the fundamental domain, $t_0$ and $t_1$ \cite{gr_sc,ce1}.
As promised, the diffractive picture captures this behaviour. For the
highest values of $k$, we used 70 partial waves in
the calculation of the diffraction constant (\ref{bluevelvet}). 
If we held the number of partial
waves fixed, the calculation would start to fail for larger values of $|kR|$.
We also include the results from the theory of geometrical orbits \cite{ce1}
for comparison. The new r\'{e}gime is at the left of the 
figure where ${\mbox{Real}}\{ka\}\ll 1$. There it can be seen that the
widths of the resonances increase logarithmically with $kR$, a result which
we generically expect for diffraction \cite{ndw1,ndw2}. In those references it
is shown that the width of the first resonance scales as $\log(d)$ and since
$d$ scales logarithmically with $a$, we find that the width of the first
resonance (when measured in units of $1/R$) scales as
$\log(\log(R/a))$, as opposed to the $\log(R/a)$ behaviour
predicted by geometric orbits \cite{creep}.
This means that in the diffractive case, 
the resonances are observable even for extremely large values of $R/a$.

In Fig.~\ref{3d_res}b
we show the results for the $A_1$ and $E$ resonances for $R/a=600$.
The agreement conforms to the discussion of the top panel,
however the increased value of $R/a$ means that none of the resonances shown
are in the geometrical r\'{e}gime ${\mbox{Real}}\{ka\}>1$
so there is no strong interference between
the weights $t_0$ and $t_1$. The $A_1$ resonances have
smaller widths because, to linear order, their zeta function 
(\ref{a_1--3d}) involves two
weights, which are in phase, while the zeta function for the $E$ resonances
(\ref{loveanddeath})
involves only one weight. A disadvantage of the three disk problem
for this study is that in the diffractive r\'{e}gime $ka\ll 1$ the s-wave term
dominates so that the two
diffraction constants $d(0)$ and $d(\pi/3)$ are very nearly equal and 
so too are the weights $t_0$ and $t_1$. The result on the spectrum is
approximately the same as if there were just one weight, a situation which
is known to lead to rather uninteresting
spectra \cite{creep,ndw1}. For this reason
we were led to study the four disk problem which we discuss next.

The four disk problem shown in Fig.~\ref{mark_4}a
has more structure than the three disk
because there are two distinct lengths in the problem; in addition to the side
length $R$, there is the diagonal length $\sqrt{2}R$.
Accordingly, we define the factor
\be \label{paristexas} 
v = {1\over \sqrt{8\sqrt{2}\pi kR}}\exp\{i(\sqrt{2}kR-3\pi/4)\},
\ee
in analogy to $u$.
In addition, there are now six distinct processes. Starting at any disk we
can go to one of the two near disks and either diffract back with $d_0=d(0)$,
diffract to the next disk with $d_1=d(\pi/2)$ or diffract diagonally
with $d_2=d(\pi/4)$. Additionally we can head diagonally across and diffract
back with $d_3=d(0)$ or to either one of the other two disks, again with 
$d_4=d(\pi/4)$. This problem has $C_{4v}$ symmetry which has four
one dimensional representations labelled $A_1$, $A_2$, $B_1$ and $B_2$ and
one two dimensional representation labelled $E$ \cite{ce2}. This system
has previously been studied semiclassically using periodic geometric
orbits \cite{fd}.

As before, we want to find the Markov graph of the problem for which we study
the dynamics in the fundamental domain which is one eighth of the full plane
and is shown in Fig.~\ref{mark_4}a. 
Starting at the half disk, we can go in one of two 
directions, which we call $A$ and $B$. We want to find all paths which start
and end at either $A$ or $B$. 
{}From $A$ we first reflect off the diagonal wall and upon returning
either diffract back which we call 0, diffract to $A'$ and then
reflect onto $A$ which we call $1$ or diffract to $B$ which we call 2.
{}From $B$ we first travel to the centre and on returning either diffract back 
which we call $3$, diffract to $A$ which we call $4$ or diffract to 
$A'$ and reflect to $A$ which we call $\bar{4}$. This is shown diagrammatically
as a Markov graph in Fig.~\ref{mark_4}b. 
Note that process 3 is a boundary orbit which
lies on a symmetry axis and can be shown to contribute only to the spectra
of representations which are not odd with respect to reflections through
that axis \cite{bent}. 

The weights corresponding to each process involve one geometric arc and one
diffraction so we find
\be \label{ladulcevita}
t_0 = d_0u \;\;\;\;\;\;\;\; t_1 = d_1u \;\;\;\;\;\;\;\; t_2 = d_2u
\;\;\;\;\;\;\;\;
t_3=d_3v \;\;\;\;\;\;\;\; t_4=t_{\bar{4}}=d_4v.
\ee
In general each one of these also has a group theoretic factor depending
on the group representation being considered. Again, we start with the
symmetric $A_1$ representation for which all the characters equal one.
Enumerating all closed loops and products of closed loops on the graph, we read
off the zeta function \cite{zet} as
\be \label{jaws}
\zeta_{A_1}^{-1} = 1-t_0-t_1-t_3-(2t_2t_4-t_0t_3-t_1t_3)
\ee
where we have used the equality between $t_4$ and $t_{\bar{4}}$.
This result involves cycles of topological lengths one and two. 
We now have contributions of length two since the graph has two nodes,
however cycles of length three and higher are absent in Eq.(\ref{jaws}). 
We again note that this is the same expression
as the cycle expansion of the 4-disk problem discussed in Ref.~\cite{ce2}
where we use use $t_{01}=t_0t_1$ and additionally invoke 
the identification between 
$\{t_0,t_1,t_3,t_2t_4,t_2t_4,t_1t_2t_4,t_0t_2t_4\}$ in our notation and
$\{t_0,t_1,t_2,t_{02},t_{12},t_{112},t_{002}\}$ in theirs.
As before, we can use this fact to read off the 
zeta functions of the other representations from Ref. \cite{ce2},
\bea
\zeta_{B_2}^{-1} & = & 1+t_0+t_1-t_3+(2t_2t_4-t_0t_3-t_1t_3)\nonumber\\
\zeta_{A_2}^{-1} & = & 1+t_0-t_1 \hspace{1.5cm}
\zeta_{B_1}^{-1} = 1-t_0+t_1 \nonumber\\
\zeta_E^{-1}     & = & 1+t_3+(t_1^2-t_0^2)+
(2t_0t_2t_4-t_0^2t_3-2t_1t_2t_4+t_1^2t_3).\label{windsofwar}
\eea

We are primarily interested in the region $ka<1$ for which the diffraction
constants are almost equal (ie the s-wave limit) and we see that the $A_2$ and
$B_1$ representations have almost total cancellation and therefore their
resonances are comparatively deep in the complex $k$ plane. These are the
representations which are odd with respect to reflections across the diagonals
of the square so process 3 does not contribute. Instead we
concentrate on the representations $A_1$, $B_2$ and $E$. 
In Fig.~\ref{4d_res}
we plot the exact positions of these three representations found using
the algorithms of Refs.\cite{trcl,stud} together with the semiclassical
approximations from Eqs.~(\ref{jaws}) and (\ref{windsofwar}) for $R/a=600$. 
In all cases, the
semiclassical predictions from the zeta functions work well although it is
interesting to note that there is a noticeable deterioration of the quality for
the resonances with large imaginary part. The irreducible representations
$A_1$ and $B_2$ have
richer spectra due to the interferences among the three basic weights. The
$E$ resonances are given by a zeta function which is dominated by the weight
$t_3$ and thereby shows the characteristic logarithmic behaviour discussed
above and observed in Fig.~\ref{3d_res}a. 
For larger values of $k$, the quadratic terms
of Eq.~(\ref{windsofwar}) become important leading to more structure in the
$E$ spectrum. This structure will eventually develop into 
the rich spectrum of scattered resonances predicted by the geometrical orbits.

In this problem, we have altogether defined 5 weights - however
since $t_2$ and $t_4$ always occur as a product in the zeta functions,
it is more precise to say there are 4 independent quantities. On the other 
hand, it
is known that the geometric orbits can be labelled with just three 
symbols \cite{ce2}. (In contrast, for the three disk problem we had only two
weights, in agreement with the two symbols needed to label geometric
periodic orbits.) This is reminiscent of the approximations of the transfer
operator based on so-called $T$ matrices \cite{bog} which lead to 
transcendantal quantisation equations like (\ref{windsofwar}) but in terms
of classical trajectories. Increasing the
dimension of the $T$ matrices induces more complicated equations in terms of
which the quantisation is more exact. That approximation is based on assuming
certain matrix elements (weights in our language) are approximately 
multiplicatively
related and so drop out of the equations - as happens exactly for diffraction.
The structural similarity between these results is presumably based on
the underlying structure of finite Markov graphs which are used by us and are
implicit in the work of \cite{bog}.

In conclusion, we have discussed a form of discontinuity which is
amenable to discussion in terms of diffraction, that of small disks. Since
the effect of a disk vanishes as the disk radius goes to zero, we must
consider disks of some fixed size. Doing so introduces a length scale in
the problem such that if $ka\gg 1$ one can use standard geometrical orbits.
However in the domain $ka\ll 1$ a qualitatively new physical
picture is necessary. The formalism we discuss here incorporates both limits
but at the price of having to include many partial waves when $ka\gg 1$.
We have tested this theory in systems consisting of three and four disks
arranged symmetrically on the plane. 
The formalism of Markov graphs and zeta functions applies equally well
to any system in which there exist objects which can be approximated as
point singularities, including point scatterers mentioned above \cite{seba}
and Aharonov-Bohm flux lines \cite{ahbohm}. These systems allow a finite
approximation based on zeta functions to give their scattering resonances
and as such are formally useful in testing the formalism. However, the
arguments developed here apply equally well to bound systems. Putting a
small disk or other singularity
inside a billiard introduces diffractive paths which 
appear in the Fourier transform of the spectrum \cite{br} 
in a characteristic way, just as with edge diffraction \cite{ps,bw}. 

The authors would like to thank Stephen Creagh and
Predrag Cvitanovi\'c for useful discussions.
A.W. would like to thank Predrag Cvitanovi\'c and the Centre for Chaos and
Turbulence Studies at the Niels Bohr Institute for hospitality and support
during his visit in August 1995. N.D.W. was supported by the European Union
Human Capital and Mobility Fund.

\begin{figure}
\caption{Left: The configuration space of the three disk problem with the
fundamental domain indicated at the top right. The arrow $A$ indicates the
unique direction in which a trajectory can leave the disk and ultimately
return and the arrow $A'$ is its mirror image. Right: The corresponding
Markov graph with a single node $A$ and the two processes which connect it
to itself.}
\label{mark_3}
\end{figure}

\begin{figure}
\caption{The same as Fig. 1 but without using the symmetry decomposition.
There are now six possible directions and consequently a six node graph.
The six short lines correspond to weights $t_0$ and the six long lines 
correspond to weights $t_1$.}
\label{mark_3a}
\end{figure}

\begin{figure}
\caption{Top: The $A_1$ three disk resonances for $R/a=60$ plotted in the
complex $k$ plane. The exact resonances are represented
as open circles, the semiclassical diffractive predictions as vertical
crosses and the semiclassical geometric predictions as diagonal crosses. 
Bottom: The
upper set of points are the $A_1$ resonances for $R/a=600$ while the lower
set are the $E$ resonances. Same symbol convention but we do not include
the geometrical orbit predictions. In both cases the wave numbers are
measured in units of $1/R$.}
\label{3d_res}
\end{figure}

\begin{figure}
\caption{Left: The configuration space of the four disk problem with the
fundamental domain indicated. The two available directions are $A$ and $B$
and $A'$ is the mirror image of $A$. Right: The corresponding Markov graph
with two nodes $A$ and $B$ and all the interconnecting processes.}
\label{mark_4}
\end{figure}

\begin{figure}
\caption{Top: The $A_1$ resonances of the four disk problem for $R/a=600$.
Middle: The $B_2$ resonances. Bottom: The $E$ resonances. (Same symbol
and unit convention as in Fig. 3.)}
\label{4d_res}
\end{figure}

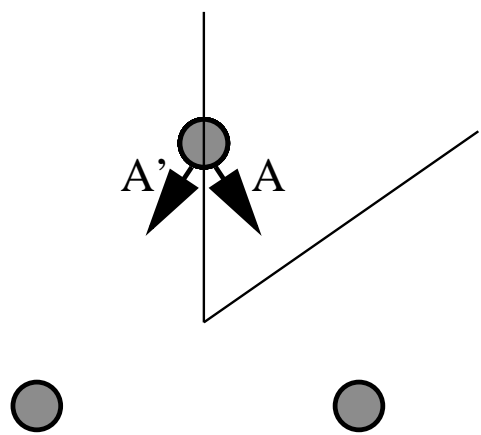
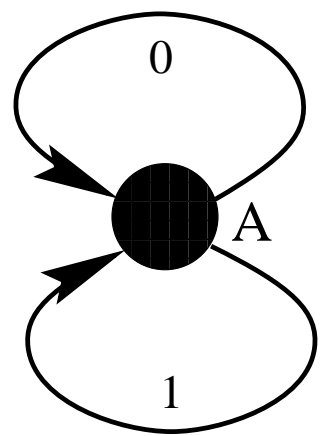
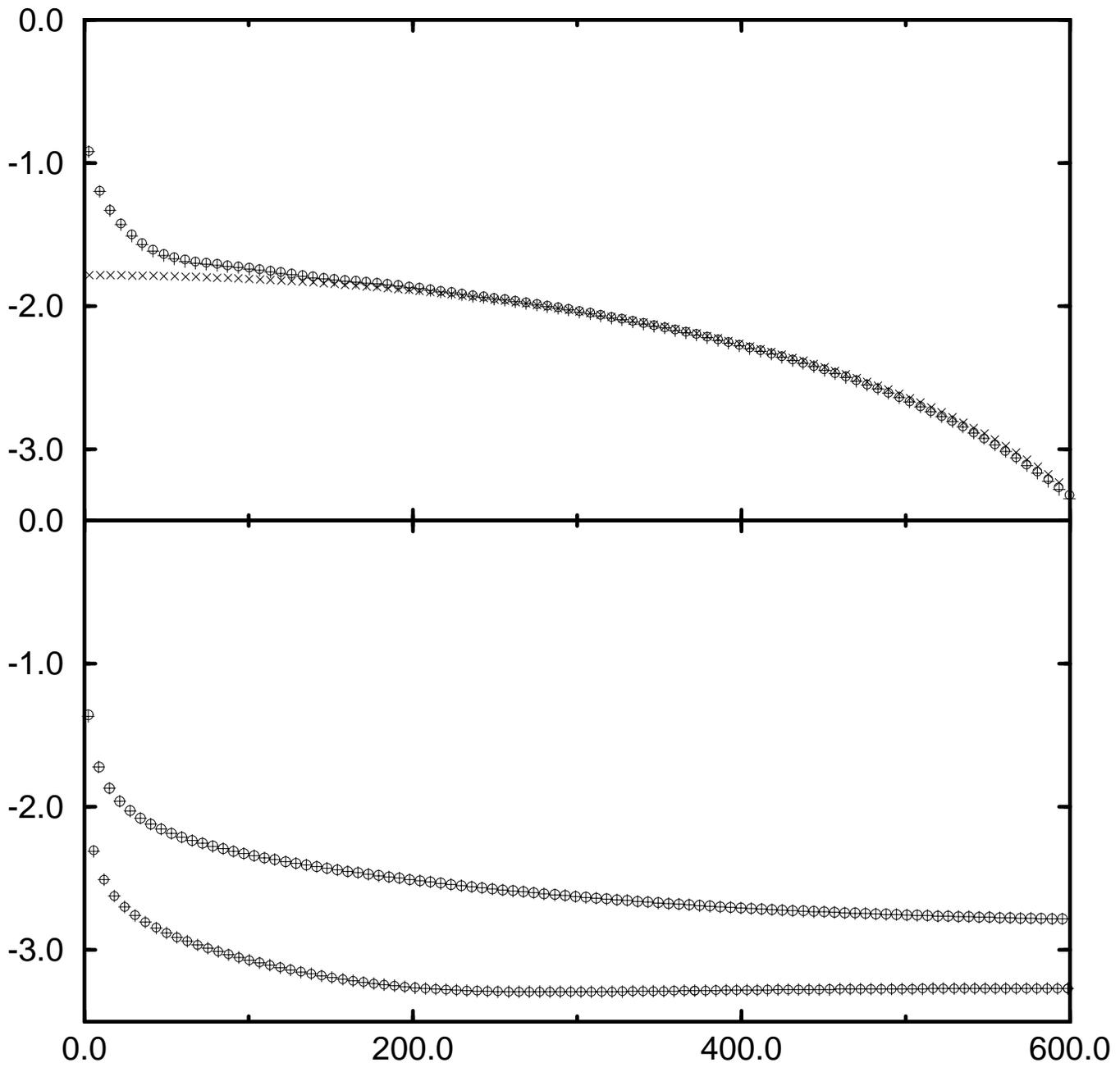
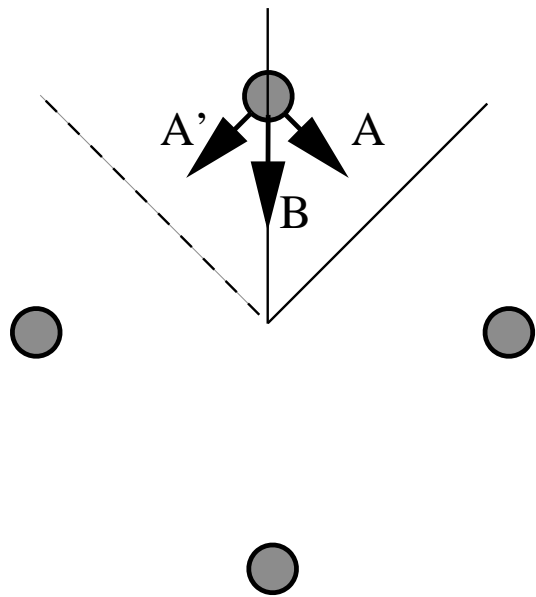
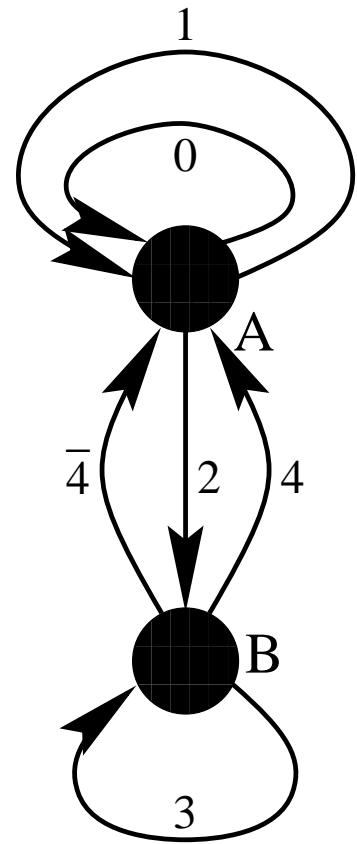
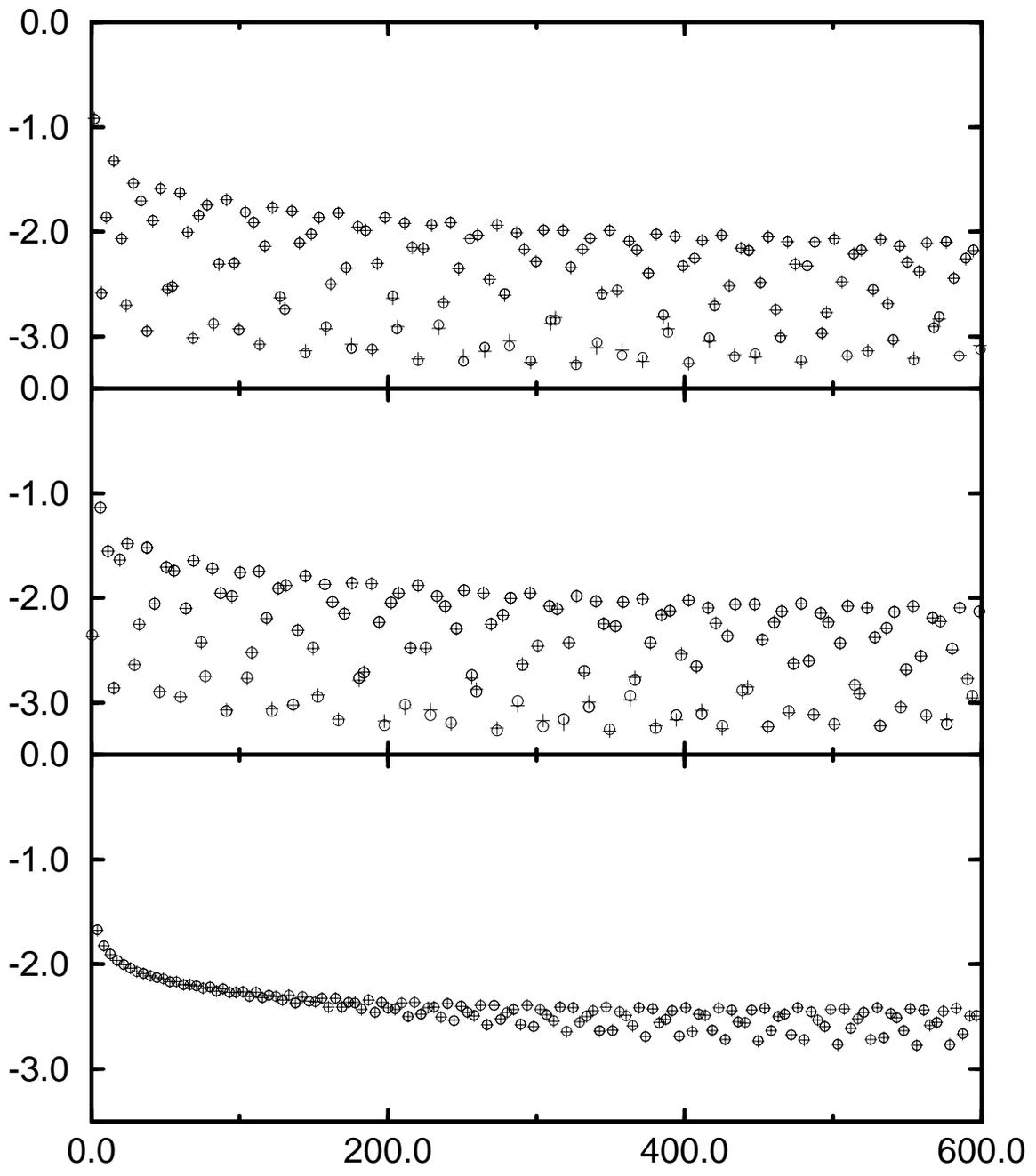
\end{document}